\documentclass[namedreferences]{SolarPhysics}
\usepackage[optionalrh]{spr-sola-addons} 
\usepackage{graphicx}
\usepackage{url}             

\newcommand{\adv}{    {\it Adv. Space Res.}}

\newcommand{\ag}{     {\it Ann. Geophys.}}

\newcommand{\apj}{    {\it Astrophys. J.}}
\newcommand{\apjl}{   {\it Astrophys. J. Lett.}}

\newcommand{\grl}{    {\it Geophys. Res. Lett.}}

\newcommand{\jgr}{    {\it J. Geophys. Res.}}

\newcommand{\pasj}{   {\it Publ. Astron. Soc. Japan}}

\newcommand{\solphys}{{\it Solar Phys.}}

\newcommand{\ssr}{    {\it Space Sci. Rev.}}

\begin{document}
\begin{article}
\begin{opening}

\title{A Challenging Solar Eruptive Event of 18 November 2003
and the Causes of the 20 November Geomagnetic Superstorm.
III.~Catastrophe of the Eruptive Filament at a Magnetic Null Point
and Formation of an Opposite-Handedness CME}

\author{A.M.~\surname{Uralov}$^{1}$~\sep
        V.V.~\surname{Grechnev}$^{1}$\sep
        G.V.~\surname{Rudenko}$^{1}$\sep
        I.I.~\surname{Myshyakov}$^{1}$\sep
        I.M.~\surname{Chertok}$^{2}$\sep
        B.P.~\surname{Filippov}$^{2}$\sep
        V.A.~\surname{Slemzin}$^{3}$}

\runningauthor{Uralov et al.} \runningtitle{Catastrophe of a
Filament and Formation of an Opposite-Handedness CME}

\institute{$^{1}$ Institute of Solar-Terrestrial Physics SB RAS,
                  Lermontov St.\ 126A, Irkutsk 664033, Russia
                  email: \url{uralov@iszf.irk.ru}\\
           $^{2}$ Pushkov Institute of Terrestrial Magnetism,
                  Ionosphere and Radio Wave Propagation (IZMIRAN),
                  Moscow, Troitsk, 142190 Russia
                  email: \url{ichertok@izmiran.ru}, \url{bfilip@izmiran.ru}\\
           $^{3}$ P.N. Lebedev Physical Institute, Leninsky Pr., 53, Moscow, 119991,
                  Russia
                  email: \url{slem@lebedev.ru} }

\date{Received ; accepted }

\begin{abstract}
Our analysis in Papers I and II (Grechnev \textit{et al.}, 2014,
Solar Phys. 289, 289 and 1279) of the 18 November 2003 solar event
responsible for the 20 November geomagnetic superstorm has
revealed a complex chain of eruptions. In particular, the eruptive
filament encountered a topological discontinuity located near the
solar disk center at a height of about 100 Mm, bifurcated, and
transformed into a large cloud, which did not leave the Sun.
Concurrently, an additional CME presumably erupted close to the
bifurcation region. The conjectures about the responsibility of
this compact CME for the superstorm and its disconnection from the
Sun are confirmed in Paper~IV (Grechnev \textit{et al.}, Solar
Phys., submitted), which concludes about its probable
spheromak-like structure. The present paper confirms the presence
of a magnetic null point near the bifurcation region and addresses
the origin of the magnetic helicity of the interplanetary magnetic
clouds and their connection to the Sun. We find that the
orientation of a magnetic dipole constituted by dimmed regions
with the opposite magnetic polarities away from the parent active
region corresponded to the direction of the axial field in the
magnetic cloud, while the pre-eruptive filament mismatched it. To
combine all of the listed findings, we come to an intrinsically
three-dimensional scheme, in which a spheromak-like eruption
originates \textit{via} the interaction of the initially
unconnected magnetic fluxes of the eruptive filament and
pre-existing ones in the corona. Through a chain of magnetic
reconnections their positive mutual helicity was transformed into
the self-helicity of the spheromak-like magnetic cloud.

\end{abstract}

\end{opening}

\section{Introduction}
 \label{S-introduction}

An extreme geomagnetic storm on 20 November 2003 was caused by the
interaction of the Earth's magnetosphere with an interplanetary
magnetic cloud (MC), whose magnetic helicity, $H_\mathrm{m}$, was
positive. The very strong magnetic field in the MC of up to
$|B|_{\max} \approx 56$~nT had a long-lasting negative $B_z$
component (up to $B_{z\, \max} \approx -46$~nT). These
circumstances were crucial in identifying the solar source for the
MC. \inlinecite{Gopal2005} and \inlinecite{Yurchyshyn2005}
definitely associated the MC with the filament eruption in the
active region (AR) 10501 (we will use henceforth the last three
digits for brevity) on 18 November. These authors considered the
direction of the axial magnetic field in the pre-eruptive filament
to correspond to the expected projection $B_z < 0$. With such a
direction of the axial field, the current helicity of the
filament, $H_\mathrm{c}$, is positive. From the condition
$\mathrm{sign} (H_\mathrm{m}) = \mathrm{sign} (H_\mathrm{c})$,
which is valid for a linear force-free magnetic field, it follows
that the magnetic helicity is also positive. Later
\inlinecite{Moestl2008} found a correspondence between the flare
reconnected magnetic flux, measured as the flare ribbon flux, and
the poloidal magnetic flux of the MC under the assumption that the
MC was a part of a magnetic flux rope with a length of 0.5--2~AU.
The studies by \inlinecite{Gopal2005},
\inlinecite{Yurchyshyn2005}, and \inlinecite{Srivastava2009} gave
the impression of an acceptable correspondence between the
conditions of the eruption in AR~501 and the parameters of the MC
observed in the Earth orbit: $H_\mathrm{m} > 0, B_z < 0$.

The study of \inlinecite{Chandra2010} changed this situation. From
the observed morphological features they found that the
large-scale magnetic field in AR~501 had a negative helicity sign.
This finding seemingly contradicted what was expected from the
magnetic helicity conservation requiring the same sign of the
magnetic helicity in the AR and MC. This circumstance has raised a
question, why the AR, which had a global negative magnetic
helicity, could expel a positive-helicity MC to the interplanetary
medium.

One possible answer was proposed by \inlinecite{Chandra2010}, who
found a localized positive helicity injection in the southern part
of AR~501 and concluded that the right handedness of the observed
MC was due to the ejection from this portion of the AR. On the
other hand, \inlinecite{Leamon2004} in their study of 12
interplanetary MCs and related solar active regions have found:
\textit{i})~a significant difference between the total twist of
the magnetic field inside active regions, $(\alpha
L)_\mathrm{AR}$, and that in the MC, $(\alpha L)_\mathrm{MC}$;
\textit{ii})~the absence of any significant sign relationship
between them. The authors used the linear force-free
approximation, $\alpha$ is a constant. The dipole scale,
$L_\mathrm{AR}$, was measured as the distance between the
centroids of the positive and negative fluxes in the magnetogram
of an AR. The magnetic field in an MC was fit with the Lundquist
magnetic model with $L_\mathrm{MC} = 2.5$~AU length. Findings
(\textit{i}) and (\textit{ii}) have compelled
\inlinecite{Leamon2004} to conclude that ``\textit{magnetic clouds
associated with active region eruptions are formed by magnetic
reconnection between these regions and their larger-scale
surroundings, rather than simple eruption of preexisting
structures in the corona or chromosphere}''.
\inlinecite{ZhangLow2003} described a similar phenomenon
analytically by using the idealized example of the
axially-symmetric reconfiguration of two twisted magnetic fluxes
from their unconnected initial state to the connected relaxed
state. They have shown that magnetic reconnection can reverse the
twist direction of a flux rope emerging into preexisting fields
under the conservation of the total relative magnetic helicity.
The complex reconnection of a flux rope with the adjacent field in
complex magnetic topology has been also described by,
\textit{e.g.}, \inlinecite{Lugaz2011},
\inlinecite{Zuccarello2012}, and \inlinecite{Masson2013}.

\citeauthor{Grechnev2008}
(\citeyear{Grechnev2008,Grechnev2011_AE,Grechnev2013_anomal}) have
found observational evidence of magnetic reconnection between the
internal field belonging to the eruptive filament and the
preexisting coronal field. This is a phenomenon of plasma
dispersal from an eruptive filament over the solar surface that is
visible as the disintegration of the filament. The whole mass of
an eruptive filament or a considerable fraction of its mass does
not leave the Sun as a part of a CME. The motion of the cool
plasma of the eruptive filament continues along new magnetic field
lines passing inside the eruptive filament and ending far on the
solar surface. Clouds of such plasma can screen the emission of
compact sources in active regions as well as the emission from
quiet solar areas. Absorption phenomena can be observed in
microwaves and also in the He~\textsc{ii} 304~\AA\ line. Events of
such a kind are associated with active region eruptions. They have
been rarely detected for observational reasons.

\begin{figure} 
  \centerline{\includegraphics[width=\textwidth]
   {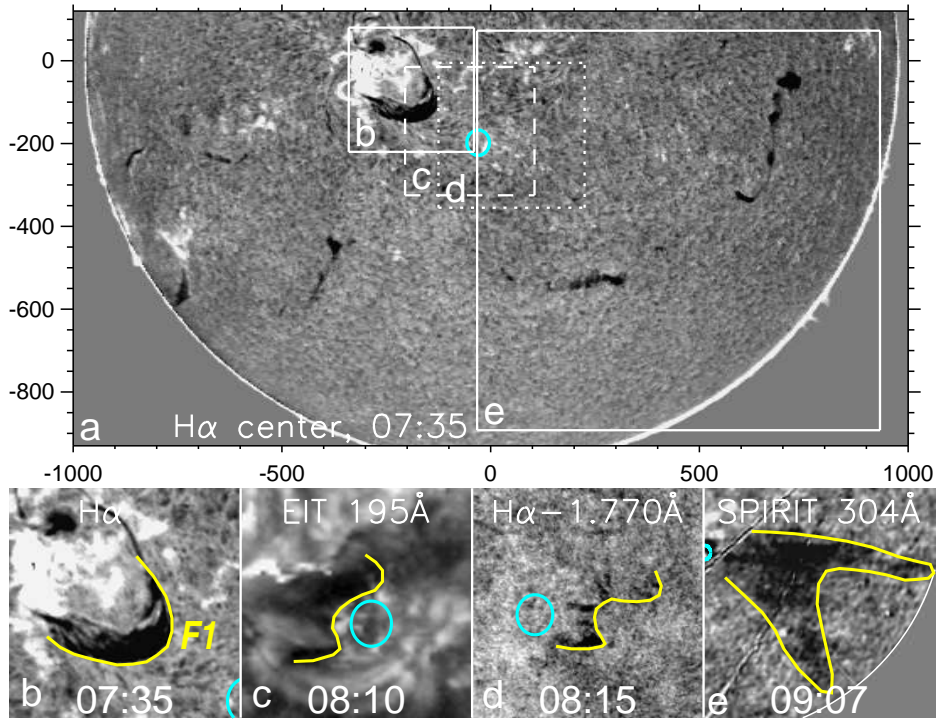}
  }
  \caption{Bifurcation of the main eruptive filament F1.
(a,b)~Pre-eruption H$\alpha$ line-center image (KSO). The frames
denote the fields of view of the four images shown in the lower
row (b--e). The axes are in arc seconds from the solar disk
center. The turquoise oval in all the images denotes the region Rb
where the filament bifurcated. The yellow curves roughly outline
the frontal edge of the filament before and during the eruption.
(c)~SOHO/EIT image in the 195~\AA\ channel. (d)~KSO H$\alpha$
image observed in the far blue wing. (e)~Y-like cloud in the
CORONAS-F/SPIRIT 304~\AA\ image.}
  \label{F-shape_transform}
  \end{figure}

The analysis of the solar geoeffective event of 18 November 2003
by \inlinecite{Grechnev2014_I} (hereafter Paper~I) and
\inlinecite{Grechnev2014_II} (hereafter Paper~II) has revealed
that the major eruption in this event, \textit{i.e.}, the eruption
of the U-shaped filament, which we call F1, from AR~501, was also
not a simple one. The eruptive filament bifurcated and transformed
into a large Y-shaped cloud, which moved from the region of
bifurcation (Rb) to the South--West across the solar disk toward
the limb. Figure~\ref{F-shape_transform} illustrates what has
happened to the eruptive filament, as shown by the H$\alpha$
images produced by the Kanzelh{\"o}he Solar Observatory (KSO), the
\textit{Extreme-ultraviolet Imaging Telescope} (EIT;
\opencite{Delab1995}), on board the \textit {Solar and
Heliospheric Observatory} (SOHO), and the
\textit{Spectroheliographic X-ray Imaging Telescope} (SPIRIT;
\inlinecite{Zhitnik2002} and \inlinecite{Slemzin2005}), on board
the \textit{Complex Orbital near-Earth Observations of Activity of
the Sun} (CORONAS-F) satellite
\cite{{OraevskySobelman2002},{Oraevsky2003}}.

The masses of the Y-like cloud and the pre-eruption filament were
similar (Paper~I); on the other hand, remnants of the filament
were not evident in the southwestern CME observed by the
\textit{Large Angle and Spectrometric Coronagraph} (LASCO) that
was previously regarded as the source of the 20 November
geomagnetic storm. The observations analyzed in Paper~I and
Paper~II suggest a possible additional eruption in the interval
from 08:07 to 08:14~UT above the bifurcation region close to the
solar disk center that could be the source of the interplanetary
MC on 20 November. These facts disfavor the simple scenario, in
which the 20 November MC is considered as a flux rope formed
directly from a structure initially associated with the
pre-eruptive filament F1 in region 501.

As mentioned, the right-handed MC produced in this event and
responsible for the superstorm had a very strong magnetic field
near Earth of up to $|B|_{\max} \approx 56$~nT and $B_{z\, \max}
\approx -46$~nT. According to \inlinecite{Moestl2008}, its
inclination to the ecliptic plane was $\theta = -(49-87)^{\circ}$,
and the magnetic flux in this plane was $0.55 \times 10^{21}$~Mx;
however, its significant part could be lost by reconnection in the
interplanetary space. In \inlinecite{Grechnev2014_IV} (hereafter
Paper~IV), we additionally find that the MC was compact, with a
size of about 0.2~AU, and had some atypical properties, such as a
wide range of proton temperatures, from $2 \times 10^4$~K to
$3\times 10^5$~K; its magnetic structure was closed, disconnected
from the Sun, and probably had a spheromak configuration.

The present paper (hereafter Paper~III) endeavors to understand
how the catastrophe of the eruptive filament could occur and
create the right-handed spheromak-like MC. Section~\ref{S-outline}
outlines the eruptive event and results of its analysis.
Section~\ref{S-helicity} analyzes the helicity in AR~501. In
Section~\ref{S-large_scale_config} we address the causes of the
bifurcation of the eruptive filament. In
Section~\ref{S-mc_formation} we try to understand how the MC could
be formed. Section~\ref{S-summary} briefly summarizes the outcome
of this study.

\section{Outline of the Event and Observational Indications}
 \label{S-outline}

\subsection{Observational Results and Suggestions}

\begin{figure} 
  \centerline{\includegraphics[width=\textwidth]
   {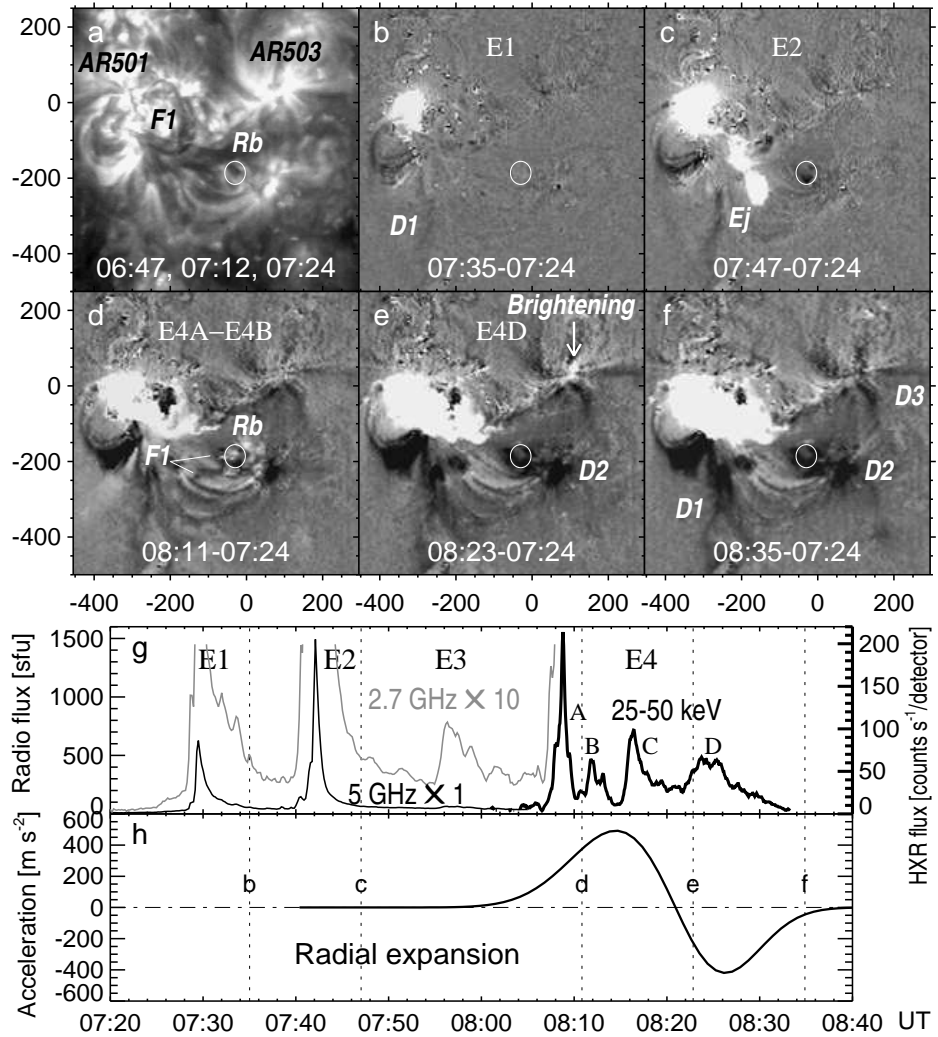}
  }
  \caption{History of the event. (a)~Pre-event situation in an averaged
EIT 195~\AA\ image, which is built with images taken at the hours
indicated at the bottom of the panel. (b--f)~Milestones in the
event in the EIT 195~\AA\ fixed-base difference images (the
differences are taken between the hours indicated at the bottom of
each panel). The solar rotation was compensated to 08:20~UT. The
axes are in arc seconds from the solar disk center. (g)~Time
profiles in microwaves at 5~GHz (thin black line), 2.7 GHz (gray
line, magnified by a factor of 10), and in HXR (thick black line).
(h)~The acceleration of the radial expansion of the invisible CME,
which presumably erupted near the solar disk center close to
region Rb. The labels E1--E4D in panels (b--f) denote the eruptive
episodes marked in the time profiles in panel (g). The vertical
dotted lines in panels (g) and (h) mark the observation times of
panels (b--f).}
  \label{F-history}
  \end{figure}

Paper~I and Paper~II have revealed a few eruptive episodes in the
complex event of 18 November 2003. Here we outline and illustrate
them using the images from EIT in
Figures~\ref{F-history}a--\ref{F-history}f and the time profiles
in Figure~\ref{F-history}g produced in hard X-rays (HXR) by the
\textit{Reuven Ramaty High-Energy Solar Spectroscopic Imager}
(RHESSI: \opencite{Lin2002}) and, for the RHESSI nighttime, in
microwaves by the US Air Force \textit{Radio Solar Telescope
Network} radiometers in San Vito.

The pre-eruptive U-shaped filament F1 rooted in AR~501 was
pointing in the plane of the sky toward a small ring-like
structure Rb (Figure~\ref{F-history}a). Active region 503 was
located North of Rb. Episode E1 presumably gave rise to a first
southeastern CME (CME1) responsible for am elongated dimming D1,
which started to develop during the interval shown in
Figure~\ref{F-history}b. A jet-like ejection, Ej, in
Figure~\ref{F-history}c (E2) triggered the motion of filament F1,
but did not produce any CME directly. The filament accelerated in
a weak episode E3 (Figure~\ref{F-history}g). The eruptive
filament, whose trajectory crossed a topological discontinuity
located at a height of about 100~Mm, collided with a coronal
structure above Rb (E4A) and bifurcated (E4B), apparently rolling
around it. In response to the interaction, brightenings
overlapping Rb appeared in Figure~\ref{F-history}d. They rotated
clockwise and, after E4B, vanished in Figure~\ref{F-history}e.
Episode E4C was presumably related to the onset of the second
southwestern CME (CME2). In Figure~\ref{F-history}e, dimmings
developed at the previous position of Rb and West of it (D2); a
central brightening (not mentioned in Papers I and II) appeared in
AR~503 (E4D), indicating its involvement in an eruption. After the
chain of eruptions E1--E4, regions D1, D2, D3, and the former Rb
region dimmed (Figure~\ref{F-history}f); the bifurcated filament
F1 apparently transformed into a Y-like `cloud' moving across the
solar disk (Figures
\ref{F-shape_transform}c--\ref{F-shape_transform}e).

One more ejection probably occurred near the solar disk center
during the E4 burst (Figures \ref{F-history}d, \ref{F-history}e).
The ejection was detected in Paper~II as a faintly visible round
feature, which expanded approximately from the position of the
bifurcation region in the images produced by the \textit{Solar
X-ray Imager} (SXI) on board the \textit{Geostationary Operational
Environmental Satellite} (GOES-12). The observed radial expansion
was coordinated with the trajectory of a drifting type IV radio
burst in the dynamic radio spectrum using the acceleration profile
in Figure~\ref{F-history}h. The final speed of this radial
expansion was $V_r \approx 100$~km~s$^{-1}$. The mass of this
ejection should be $\ll 5 \times 10^{15}$~g (see Paper~I). The
properties of the presumable ejection are very different from
those of CME2 ejected at the same time, but appear to match the
expectations for a source of the 20 November superstorm. This
low-mass, weakly expanding ejection presumably moved along the
Sun--Earth line, and therefore its meager Thomson-scattered light
was insufficient to be detected by the LASCO coronagraphs. On the
other hand, neither CME1 nor CME2 seem to be a promising candidate
to be the source of the superstorm, being able to produce, at
most, a glancing blow on the Earth's magnetosphere (Paper~II).

According to three reconstructions of the MC responsible for the
geomagnetic superstorm \cite{Yurchyshyn2005, Moestl2008, Lui2011},
its dimensions in the ecliptic plane were $< 0.3$~AU. This size
corresponds to an expansion angle of $< 17^{\circ}$, which is
similar to that of the presumable CME (Paper~II). To meet the
Earth, the MC expanding in such a narrow angle should be ejected
close to the solar disk center. Such a weak expansion is favored,
if the MC is disconnected from the Sun. These speculations will be
supported in Paper~IV.

The observational suggestions and the listed conjectures indicate
that the ejection responsible for the superstorm probably
originated during the E4 burst. The formation of the ejection
started at E4A at a height of about 100~Mm ($\approx
0.14R_{\odot}$) and was completed by the end of E4D. The latter is
supported by the appearance of the central brightening in region
AR~503 (Figure~\ref{F-history}e). A possible height, at which the
formation was completed, is of the order of $V_\mathrm{transit}
\times \Delta t_{\mathrm{E4}} \approx 1.7R_{\odot}$, with $\Delta
t_{\mathrm{E4}}$ being the duration of burst E4, and
$V_\mathrm{transit} \approx 800-900$~km~s$^{-1}$ being an average
Sun--Earth transit speed of the ICME (estimated from the decrease
of the Dst index on 20 November). This height can be overestimated
by a factor of 2--3 due to the uncertainties in the velocity as
the CME can accelerate as it leaves the Sun and decelerate during
its transit away from the Sun. A probable height, at which the
formation was completed, is therefore between $0.6R_{\odot}$ and
$1.7R_{\odot}$.

\begin{figure} 
  \centerline{\includegraphics[width=\textwidth]
   {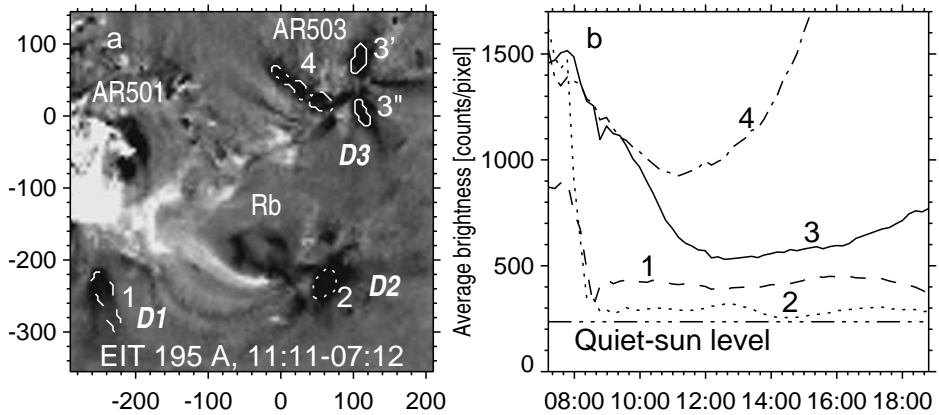}
  }
  \caption{Dimming regions near the solar disk center well
after the event. (a)~The configuration of dimmings in the EIT
195~\AA\ difference image. Five significantly dark and long-lived
dimmed regions are selected using contours of different line
styles. The solar rotation is compensated to 08:20~UT. The axes
are in arc seconds from the solar disk center. (b)~Time profiles
in separate dimmed regions. Labels 1 -- 4 denote the corresponding
regions in panel (a); the styles of the lines in panel (b) and the
corresponding contours in panel (a) are the same. The time profile
3 is averaged over regions $3^{\prime}$ and $3^{\prime \prime}$.
The horizontal line presents the average quiet-sun intensity level
in the EIT 195~\AA\ images, two of which were used for the
difference in panel (a).}
  \label{F-dimming}
  \end{figure}

It is difficult to expect to have direct observations of the
processes, which occurred at the heights previously estimated
close to the solar disk center. Therefore, we are forced to
involve indirect observational indications and calculations. We
use the observational indication provided by the evolution of the
dimming regions D1--D3 in Figure~\ref{F-dimming}. Their
configuration is visible in the EIT 195~\AA\ difference image
shown in Figure~\ref{F-dimming}a. The time profiles of the average
brightness in selected regions having the lowest intensity and
longest lifetime are presented in Figure~\ref{F-dimming}b. Unlike
\inlinecite{Attrill2007}, who defined a dimmed region as the
brightness decrease below the quiet-sun brightness level (the
horizontal line in Figure~\ref{F-dimming}b), we select the
dimmings simply by a decrease of the brightness with respect to
the pre-event level. Our different selection criterion indicates
that we are considering a different kind of phenomenon.

Most likely, the major cause of a dimming (in our definition) is
the density decrease in the coronal structures due to their
expansion. The brightness in the extreme-ultraviolet (EUV) and
soft X-ray images is proportional to the column emission measure.
The brightness, $I$, of an expanding closed coronal structure of a
linear size, $L$, and area, $A \propto L^2$, filled with a fixed
total number of emitting particles, $N_0$, should be $I \propto
EM/A \propto n^2L = (N_0/V)^2L \propto 1/L^5$. Thus, the expansion
alone should result in a considerable brightness decrease and a
strong pressure gradient, which causes a secondary subsonic plasma
outflow in the footprint regions of a CME \cite{HarraSterling2001,
Imada2007, Jin2009}. On the one hand, the outflow is responsible
for the redistribution of the coronal plasma from footprints into
the expanding volume. On the other hand, due to the subsequent
plasma supply from the chromosphere-to-corona transition region,
the outflow probably becomes the major factor to recover the
brightness in the dimmed regions. This simple consideration also
explains why the development of dimming is often observed to start
before the eruption, when coronal structures gradually expand
during the initiation phase.

The time profiles in Figure~\ref{F-dimming}b show that different
parts of the star-like dimming D3 recovered considerably faster
than the long-lived dimmings D1 and D2. The analysis of the
magnetic connectivity shows that the plage region, where dimming
D2 occurred, was connected to AR~503 associated with dimming D3.
In addition, a fan of long diverging field lines rooted in the
plage region connected it to remote regions far to the
South--West. Possibly, D2 shared a footprint of CME2 and had
therefore the time profile typical of long-lived core dimmings.
The pronouncedly shorter lifetimes of regions $3^{\prime}$,
$3^{\prime \prime}$, and 4 of the atypical dimming D3 hints at its
involvement into an eruption, in which a structure disconnected
from the Sun developed. The disconnection produced a stretched
magnetic loop. Its subsequent evolution is determined by the
relation between the magnetic tension, which tends to shrink the
loop, and the opposite influence of the plasma pressure and the
plasma outflow responsible for the dimming in its bases. The
shrinking duration for the loop with a length of $L \sim
1.7R_\odot \approx 10^3$~Mm is presumably comparable to the time
required to stop the plasma outflow, $\Delta t \sim L/C_\mathrm{S}
\sim 10^4$~s ($C_\mathrm{S} \sim 10^2$~km~s$^{-1}$ is a mean sound
speed of the plasma outflow originating from the low corona up to
the chromosphere-to-corona transition region). This estimate is of
the order of the observed formation time of dimming D3.

\begin{figure} 
  \centerline{\includegraphics[width=0.85\textwidth]
   {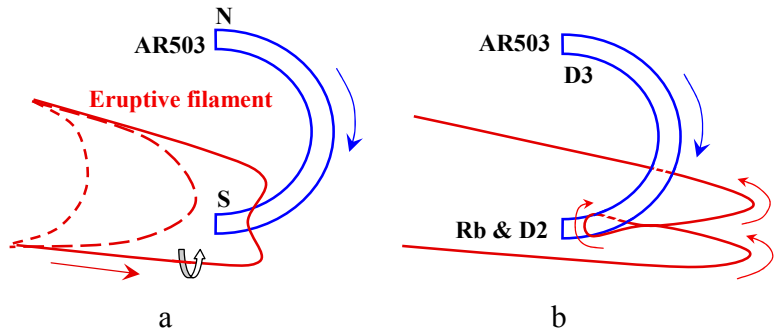}
  }
  \caption{Presumable interaction between the eruptive
filament (red) and a static coronal structure (blue) suggested by
the observational hints. The static structure connects the
N-polarity of AR~503 with an S-polarity plage region and region of
bifurcation Rb. The arrows indicate the directions of the axial
magnetic fields. The curled arrow indicates the direction of the
poloidal flux in the filament. (a)~The onset of the interaction.
(b)~Subsequent expansion of the eruptive filament. The interaction
results in the development of dimmings D3, D2, and at the former
position of Rb.}
  \label{F-suggestion}
  \end{figure}

Figure~\ref{F-suggestion} summarizes the listed observational
indications of a possible interaction between the eruptive
filament (red) and a large static coronal structure (blue). One of
its ends is rooted in AR~503, and the opposite-polarity basis
corresponds to a plage region and region of bifurcation, Rb. The
onset of the interaction and bifurcation of the filament in shown
in Figure~\ref{F-suggestion}a. In a subsequent expansion of the
eruptive filament in Figure~\ref{F-suggestion}b, its top brakes at
the blue structure, while its lateral portions stretch out in a
Y-like form. In response to the interaction, dimmings develop in
the plage region (D2) and Rb as well as dimming D3 at the opposite
end of the blue structure rooted in AR~503. Most likely, the
eruptive filament was far from AR~503 (\textit{cf.} Figures
\ref{F-history} and \ref{F-dimming}), which nevertheless appears
to be implicated. We shall use this presumable scheme as a hint in
our considerations.

\subsection{Analysis Techniques}

A major step in our study is the analysis of the coronal magnetic
field \textit{via} the extrapolation of photospheric magnetograms.
For AR~501 we have a vector magnetogram that makes possible a
non-linear force-free (NLFF) field extrapolation within its field
of view, $315^{\prime \prime} \times 315^{\prime \prime}$.
However, the region of interest, including Rb, D2, and D3 in
Figure~\ref{F-dimming}a, is far West from the vector magnetogram.
The only possible way to analyze a larger region is a potential
extrapolation of the full-disk SOHO/MDI magnetograms
\cite{Scherrer1995}. The potential approximation is usually
considered to be insufficiently accurate to describe realistic
configurations under the presence of significant electric
currents. Nevertheless, in Paper~I we have used it successfully to
visualize the filament, its height, and the topological
discontinuity, whose presence accounted for the apparent
disintegration of the eruptive filament. Let us try to check, how
realistic the results of the potential extrapolation are by
comparing them with real coronal loops in EUV images.

\begin{figure} 
  \centerline{\includegraphics[width=\textwidth]
   {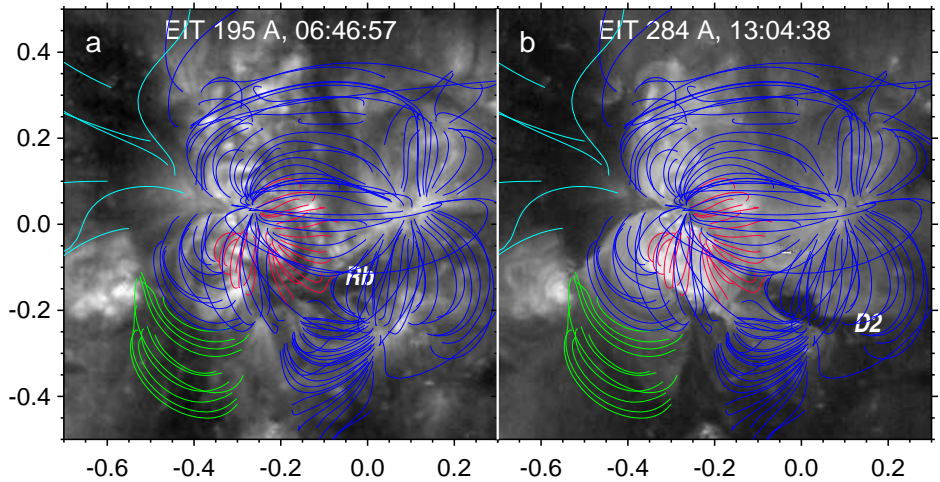}
  }
  \caption{Extrapolated field lines overlaid on: (a)~a pre-event
EIT 195~\AA\ image and (b)~a post-event EIT 284~\AA\ image. All
data were taken on 18 November. The MDI magnetogram at 09:35~UT
was used for the potential extrapolation. The EIT images are
corrected for the solar rotation to this time. The axes are in
solar radii from the solar disk center.}
  \label{F-f_lines_center}
  \end{figure}

Figure~\ref{F-f_lines_center} compares the field lines computed in
the potential approximation, using the method and package of
\inlinecite{RudenkoGrechnev1999} and \inlinecite{Rudenko2001} for
the full-disk SOHO/MDI magnetogram observed on 18 November at
09:35~UT. The spatial resolution is determined by 90 spherical
harmonics. The starting points for all of these field lines were
chosen manually, and their density does not correspond to the real
magnetic field strength distribution. The correspondence between
the computed field lines and the visible coronal loops is what is
relevant to us.

Figure~\ref{F-f_lines_center}a presents the EIT 195~\AA\ image
observed before the event in gray scale overlaid with a few sets
of computed magnetic field lines. The same sets of the field lines
are overlaid on the post-event EIT 284~\AA\ image in
Figure~\ref{F-f_lines_center}b. The red lines embrace the
pre-eruptive filament F1. The light-blue lines correspond to open
magnetic field lines rooted in the coronal hole East of AR~501.
The green lines cover a southern filament channel; the
corresponding coronal loops appeared after the event in
Figure~\ref{F-f_lines_center}b. All other structures are traced by
the blue lines.

Although a one-to-one correspondence is not observed, as expected,
the figure shows an acceptable similarity of the computed blue
field lines to the structures in the EUV images. The pre-event EIT
195~\AA\ image in Figure~\ref{F-f_lines_center}a presents rather
short loops reaching small heights. The computed field lines
correspond to the loops diverging from AR~503 and from the plage
region slightly South--West of Rb. The overall correspondence in
shapes and directions between the computed and real structures is
observed to the South--East and North--East of AR~501. The red
field lines above the filament correspond to magnetic structures,
which are expected to be seen but are not visible. These lines
fairly correspond to the post-eruption arcade to the South of
AR~501 in Figure~\ref{F-f_lines_center}b. The light-blue open
field lines are not expected to be visible.

The post-event EIT 284~\AA\ image in
Figure~\ref{F-f_lines_center}b shows somewhat higher loops. For
example, connections between AR~501 and AR~503 become detectable
here. In addition to the mentioned features, note that region Rb
is not seen here. Dimming D2 resembling a transient coronal hole
has decreased in size with respect to its appearance in
Figure~\ref{F-dimming}.

\begin{figure} 
  \centerline{\includegraphics[width=0.6\textwidth]
   {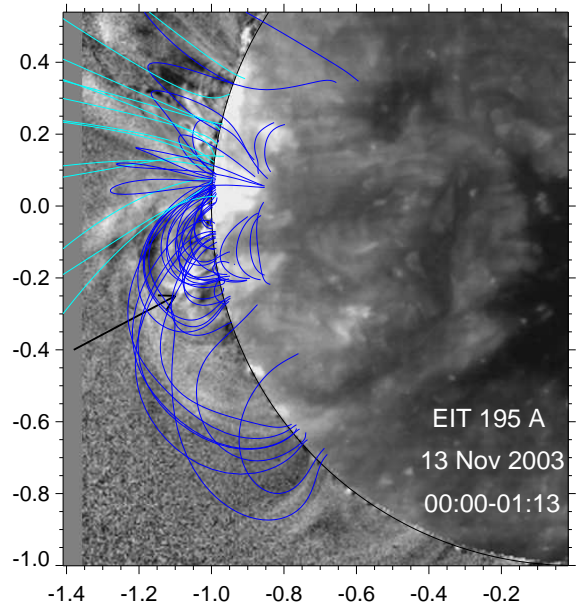}
  }
  \caption{Comparison of the field lines computed in the potential
approximation for the MDI magnetogram produced at 09:35~UT on 18
November with a synthetic EUV image composed from EIT 195~\AA\
images actually observed on 13 November. The black arrow indicates
the position of the topological discontinuity as it was on 18
November. The axes are in solar radii from the solar disk center.}
  \label{F-f_lines_limb}
  \end{figure}

We cannot observe in these images the structures, which reach
still larger heights, because of their lower brightness (due to a
lower plasma density) against the brighter lower loops. They can
only be detected against the sky on the limb. However, the on-limb
magnetogram is very uncertain. We therefore have computed the
field lines extrapolated from the magnetogram observed on 18
November at 09:35~UT and rotated them to 13 November. This day was
characterized by ongoing activity on the East limb. We have built
a composed off-limb EUV image from the ratios of several EIT
195~\AA\ images observed from 00:00~UT to 05:00~UT, bypassing the
intervals of activity. The on-disk part is an average of six EIT
195~\AA\ images observed from 00:00~UT to 01:13~UT, in which the
solar rotation was compensated. The result is presented in
Figure~\ref{F-f_lines_limb}. The light-blue lines correspond to
the open field. In spite of the five-day difference and active
conditions, there is an overall correspondence between the
calculated field lines and the off-limb structures.

In summary, Figures \ref{F-f_lines_center} and
\ref{F-f_lines_limb} demonstrate that the extrapolation in the
potential approximation presents a more or less realistic picture.
Furthermore, we observe that despite the ongoing activity, the
distribution of coronal magnetic field remained relatively stable.

\section{The Magnetic Helicity Issue}
 \label{S-helicity}

\subsection{Suggestion of Chandra \textit{et al.} (2010) in Terms of a Simple Eruption Preserving Helicity}
\label{S-simple_eruption}

\inlinecite{Chandra2010} computed the maps of the magnetic
helicity flux density injection and, in spite of the mixed
helicity signs, showed the existence of a localized positive
helicity injection in the southern part of AR~501. They also
concluded that the positive helicity was ejected from this portion
of the AR leading to the observed positive-helicity MC. The
accumulated positive helicity, as \inlinecite{Chandra2010}
considered, was concentrated in one of the portions of the
pre-eruptive filaments, while the helicity of other portions was
negative and corresponded to the sign of the global helicity of
AR~501.

However, the idea of \inlinecite{Chandra2010} to account for the
positive helicity of the near-Earth MC does not seem to be
convincing. Firstly, the helicity injection rate measured for some
part of an active region during a rather short time interval does
not provide information on the total helicity accumulated during
the preceding evolution. Secondly, the fresh idea of the authors
about the presence of the opposite-helicity portions in the body
of the pre-eruptive filament, in our opinion, does not have a
convincing visual support. Finally, \inlinecite{Chandra2010}
pointed out that the injection rate of the positive helicity flux
was not high, so that such an injection was able to accumulate the
helicity appropriate for the MC in, at least, six days.

\subsection{Absence of Positive Helicity in the Source of the Major
Eruption}
 \label{S-helicity_absence}

\inlinecite{Chandra2010} used the orientation of the filament
barbs as a morphological indication of a section with a positive
twist in the pre-eruptive filament (Figures
\ref{F-shape_transform}a and \ref{F-shape_transform}b). These
authors found a filament segment consisting of two sections, which
had the same direction of the central axis, but opposite twists.
The authors indicated, however, that this determination was
relatively ambiguous. In our opinion, the identification of the
filament barbs and their usage as a morphological indication is
questionable for such a broad filament. According to Paper~I, the
filament was tilted by about $60^{\circ}$ to the solar surface,
and features as the filament barbs could correspond to different
parts located at different heights (see Figures 14 and 17 from
Paper~I).

To confirm the suggestions inferred from the apparent filament
barbs, \inlinecite{Chandra2010} invoked different methods,
\textit{i.e.}, magnetic field extrapolation based on the linear
(constant $\alpha$) force-free field approximation, and the
computation of the magnetic helicity flux density injection. Using
a magnetogram before the eruptions (at 06:23~UT), they found a
dominant negative $\alpha$ by matching the shapes of the computed
field lines to that of the observed coronal loops. This finding
confirms their conclusion that the large-scale magnetic field of
AR~501 had a negative helicity sign. The authors also compared the
computed field lines with the post-flare arcade loops in the TRACE
171~\AA\ full-resolution image \cite{Handy1999} at 09:43~UT and
found that the loops could be modeled by using any small $\alpha$
value --- positive, or negative, or null. They also found that,
despite the predominance in the AR~501 of negative $\alpha$
values, corresponding to the negative global magnetic helicity, a
small $\alpha$ tended to be positive in the southeastern part of
the AR. The authors also computed an ongoing injection of the
positive-helicity flux in this part of AR~501 and conjectured that
this localized injection could be sufficient to make the magnetic
helicity positive in this area, but its total value remained
uncertain.

\inlinecite{Yurchyshyn2005} modeled the post-flare loops visible
in a half-resolution 195~\AA\ image obtained with EIT at 09:36~UT
and found that the best value of $\alpha$ was slightly positive
along the total length of the visible arcade. This led them to a
conclusion that the AR had a global positive magnetic helicity.
The authors used the images obtained at the late post-flare stage,
when the non-potentiality of the post-flare loops was least
pronounced being, therefore, difficult to observe. On the other
hand, at the earlier stages of a flare, more favorable in this
respect, the loops are usually unresolved because of their small
size. We have succeeded in solving this problem.

\begin{figure} 
  \centerline{\includegraphics[width=\textwidth]
   {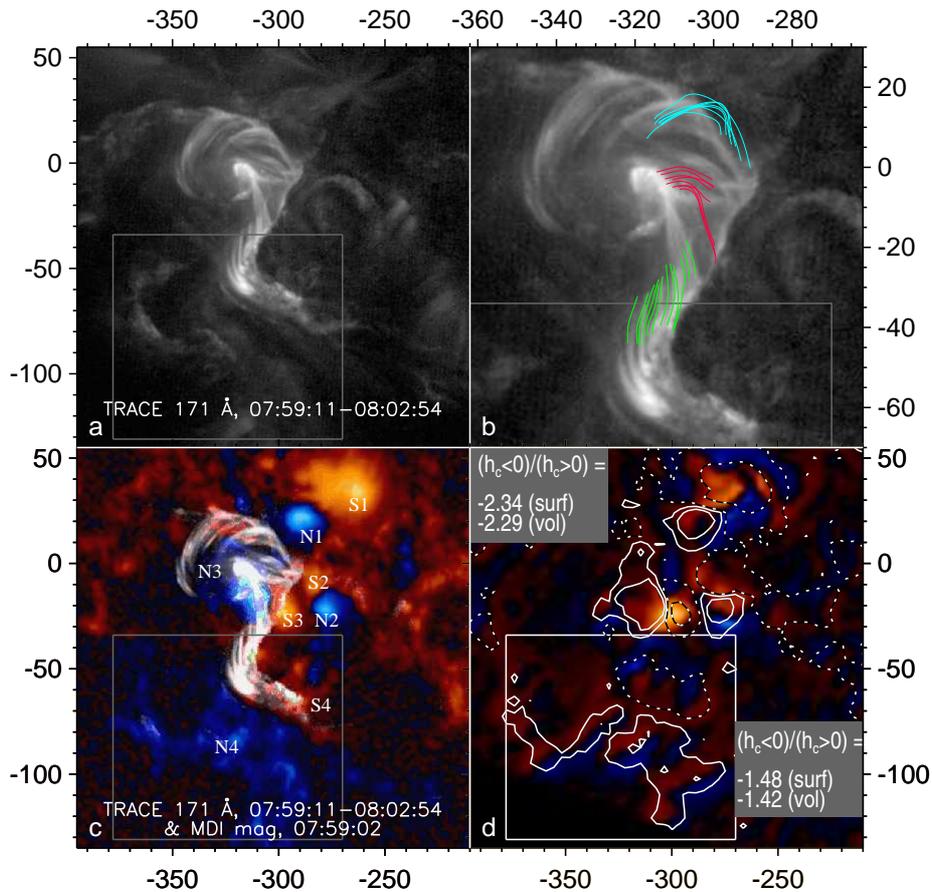}
  }
  \caption{The magnetic helicity sign issue. (a)~Inverse S-shaped arcade in a TRACE
171~\AA\ composite image. The region analyzed by Chandra
\textit{et al.} (2010) is framed in all panels. (b)~Enlarged image
of the arcade overlaid with magnetic field lines calculated from
the NLFF extrapolation. (c)~Combination of the TRACE image in
panel (a) with an MDI magnetogram (N-polarity blue, S-polarity
red, numbering according to Chandra \textit{et al.}, 2010).
(d)~Current helicity map computed from the NLFF extrapolation.
Each pixel of this map is the photospheric base of a vertical
column as high as 135~Mm. The red- (blue-) shaded regions
correspond to the values of negative (positive) current helicity
within the volume of this column. The contours outline the
negative (dotted, $-(120, 600)$~G) and positive (continuous,
$+(30, 250)$~G) $B_z$ polarities in the BBSO magnetogram used in
the computations. The top left gray box presents the ratio of the
total negative to positive current helicity at the photospheric
layer (surf) and in the coronal volume (vol) for the whole region.
The bottom right gray box is related to the framed region. The
axes are in arc seconds from the solar disk center.}
  \label{F-helicity}
  \end{figure}

The loops of a compact arcade visible in the TRACE images in
Figures \ref{F-helicity}a--\ref{F-helicity}c (not analyzed by the
aforementioned authors) are skewed counter-clockwise in sunspot N3
and the region around it, indicating negative helicity. The
skewness of the loops in region S4 considered by
\inlinecite{Chandra2010} seems to be the same, as expected for
both ends of the same loops, connecting these regions in Figures
\ref{F-helicity}a--\ref{F-helicity}c. The whole flare arcade has
an inverse S-shape, which also corresponds to negative helicity.

To verify this suggestion, we have computed the current helicity,
$h_\mathrm{c} = \mathbf{B} \cdot (\nabla \times \mathbf{B})$, from
a vector magnetogram of AR~501 observed at \textit{Big Bear Solar
Observatory} (BBSO) on 18 November at 20:26~UT (courtesy
V.~Yurchyshyn). The field of view of the magnetogram, of
$315^{\prime \prime} \times 315^{\prime \prime}$, was centered
near the N1 sunspot (about $[-300^{\prime \prime}, +20^{\prime
\prime}]$ at 08:00~UT) and covered the active region and its
vicinity, but did not reach the region of bifurcation. The
line-of-sight component in the BBSO magnetogram was considerably
saturated in sunspots S1, N3, and S3. Our computations were
carried out in a spherical box with a photospheric base of
$22^{\circ} \times 22^{\circ}$ and a height of 135~Mm from both
the raw magnetogram and a saturation-corrected one by using a
temporally close SOHO/MDI magnetogram. We used the method of the
NLFF extrapolation \cite{RudenkoMyshyakov2009} to compute two
kinds of the current helicity maps. The first kind of map, the
photospheric $h_\mathrm{c}$ density map, presents the distribution
of $h_\mathrm{c}$ at the photospheric layer. The second kind of
map is the column density distribution of $h_\mathrm{c}$. Each
pixel of such a map is the photospheric base of a vertical column,
whose height is 135~Mm, and presents the total $h_\mathrm{c}$ over
all cells constituting this column. Comparison of the maps of the
two kinds show that they are not very different, because the major
contribution to the total current helicity is due to the lowest
layers closest to the photosphere.

Each attempt has resulted in a significant excess of the negative
helicity in both the whole active region and in the framed area
considered by \inlinecite{Chandra2010}, but with somewhat
different quantitative results. The saturation-corrected current
helicity map is shown in Figure~\ref{F-helicity}d. The ratios of
the total negative to positive helicity for both the photospheric
layer and for the total volume are specified in the gray boxes.
The top left box presents the ratios for the whole region, the
bottom right box is related to the framed region. Indeed, the
excess of the total negative helicity is less in the framed
region, in agreement with the ongoing injection of the positive
helicity found by \inlinecite{Chandra2010}, but still insufficient
to make the total helicity in this region positive. The absence of
positive magnetic helicity in the localized area of AR~501 also
indicates the absence of segments with the corresponding
handedness in the body of the main filament F1. Note also that the
central part of the inverse S-shaped structure in Figures
\ref{F-helicity}a and \ref{F-helicity}b passes between sunspots N3
and S3, where the negative current helicity is the largest in the
whole AR~501.

To make the situation clearer, we have additionally calculated
magnetic field lines from the same NLFF extrapolation, which was
used in the computation of the current helicity. It is difficult
to reach a perfect correspondence between the field lines and the
observed loops, because the rapid evolution of AR~501 during the
12.5 hours separating the EUV image and the vector magnetogram
could imply a change in the small-scale features, while the strong
saturation of the vector magnetogram (and, to a lesser degree, its
limited field of view) could considerably affect long loops. For
example, we have not succeeded to reproduce the long loops in the
North--East part of the arcade; however, its negative helicity is
undoubted. Nevertheless, three sets of field lines with $\alpha <
0$ overlaid on the enlarged image of the arcade in
Figure~\ref{F-helicity}b more or less correspond to the actually
observed loops, in particular, the green lines overlapping the
region of the questionable handedness.

All of these circumstances show that the attempts to reconcile the
handedness of the magnetic cloud and active region 501 are not
promising. The question how the right-handed MC could be formed in
the eruption of the left-handed filament remains unanswered.

\subsection{Eruptions with Mismatching Helicity Sign}
 \label{S-mismatching_eruptions}

The scenario proposed by \inlinecite{Chandra2010} corresponds to
the concept of a simple eruption, when the internal magnetic
helicity of a pre-eruptive structure, \textit{i.e.}, its
self-helicity (or twist helicity), $H_\mathrm{m}^\mathrm{self}$,
determines the helicity sign of the interplanetary MC. Probably,
this situation is typical of CMEs associated with eruptions of
quiescent filaments outside active regions. However, if a filament
(flux rope) erupts from an active region, then, under certain
circumstances, the helicity of the MC can have a different origin.
This scenario implies the interaction and reconnection between
magnetic fields of the eruptive structure with coronal magnetic
fields surrounding the parent AR. In such a case, a new magnetic
structure is formed, which is the progenitor of the future
interplanetary MC. Its helicity is determined by the sum of
$H_\mathrm{m}^\mathrm{self}$ and the mutual (or linkage) helicity,
$H_\mathrm{m}^\mathrm{mut}$, of the interacting magnetic fluxes.
\textit{Via} magnetic reconnection, the mutual helicity is
transformed into the self-helicity of the MC. Depending on the
sign of $H_\mathrm{m}^\mathrm{mut}$ and the value of the
$H_\mathrm{m}^\mathrm{mut}/H_\mathrm{m}^\mathrm{self}$ ratio, the
MC helicity can be different from the pre-eruptive structure not
only in the value, but also in the sign. Such eruptions with
mismatching helicity were suggested in the study by
\inlinecite{Leamon2004}; this scenario is also supported by the
observations of the apparent disintegration of eruptive filaments
mentioned in Section~\ref{S-introduction}.

To understand what could happen on the Sun on 18 November, we
first analyze the configuration of the magnetic field on spatial
scales considerably larger than the size of the vector magnetogram
in Figure~\ref{F-helicity}. This is possible by using full-disk
magnetograms, of which line-of sight measurements (\textit{e.g.},
by SOHO/MDI) are only available for 2003. Thus, we had to use for
this purpose MDI magnetograms and the potential field
extrapolation.

\section{Large-Scale Coronal Magnetic Configuration around the
Left-Handed Eruptive Filament}
 \label{S-large_scale_config}

\begin{figure}    
   \centerline{\includegraphics[width=\textwidth]
        {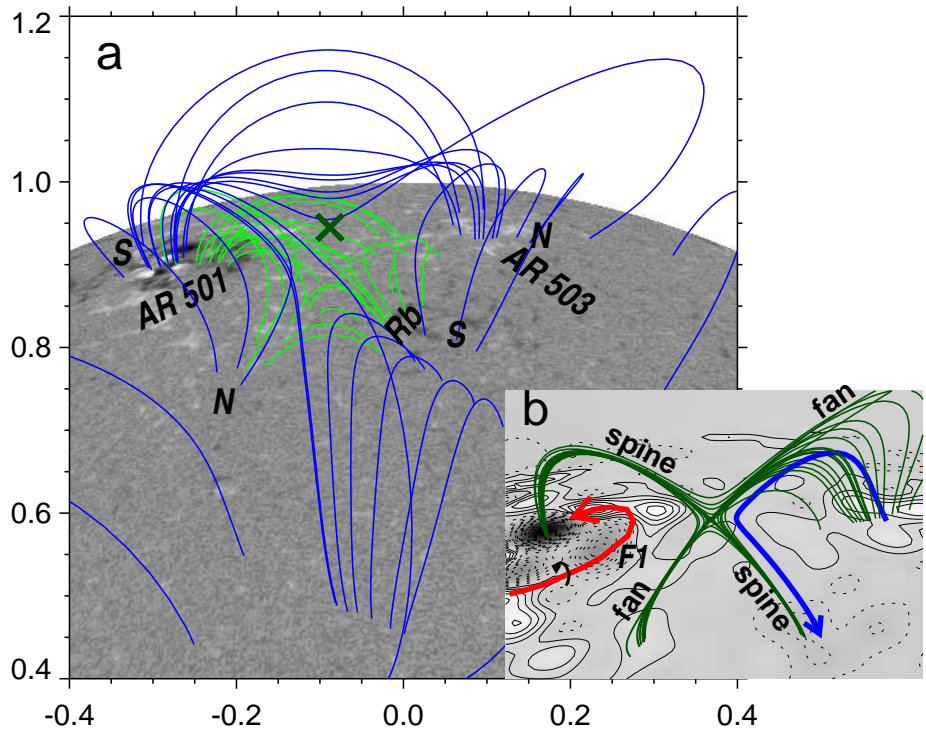}
              }
    \caption{Side view of the coronal magnetic configuration around
the pre-eruptive filament computed from the SOHO/MDI magnetogram
of 18 November, 09:35~UT. (a)~Field lines in the large-scale
environment of the magnetic null point denoted with the dark-green
slanted cross. (b)~A selected set of field lines passing close to
the magnetic null point shown in the same scale as in panel (a).
The null point at their apparent intersection corresponds to the
dark-green slanted cross in panel (a). The thick red arrow
represents the pre-eruptive filament and the direction of its
axial field. The small black circular arrow around it indicates
its negative handedness. The blue arrow represents the
connectivity of the west N--S dipole, which is a part of the
S--N--S--N quadrupole. The relative position of the red and blue
arrows suggests a positive handedness. The contour levels of the
radial magnetic component are symmetric relative to zero in steps
of 31~G: dotted $-(15, 46, ... 728)$~G, continuous $+(15, 46, ...
139)$~G. The axes are in solar radii from the solar disk center.
              }
 \label{F-field_lines}
   \end{figure}

The coronal field in a magnetic complex consisting of regions 501,
503, and their environment was computed in the potential
approximation using the mentioned package
\cite{RudenkoGrechnev1999,Rudenko2001} for the full-disk SOHO/MDI
magnetograms of 18 November observed at 06:23, 07:59, and
09:35~UT. The computations used 90 spherical harmonics. The field
line distribution computed from each of the magnetograms was
similar; we mainly use here the magnetogram at 09:35~UT.
Figures~\ref{F-field_lines}--\ref{F-fil_motion} present the
coronal magnetic configuration. The S--N--S--N quadrupole in
Figures \ref{F-field_lines}, \ref{F-top_view}, and
\ref{F-fil_motion} is its basis. Figure~\ref{F-top_view} shows
that the major eastern S-polarity sunspot has a large excess of
negative magnetic flux which is unbalanced in this quadrupole. A
considerable portion of this flux is connected to remote sites on
the solar surface (Figure~\ref{F-fil_motion}) separated from
AR~501 by the global polarity inversion line.

\subsection{Magnetic Null Point and Region of Bifurcation of the Eruptive Filament}
 \label{S-null_and_bifurcation_region}

A topological particularity of the coronal configuration is a
magnetic null point located at a height of about 100~Mm above the
photosphere. This is the only null point associated to large-scale
magnetic fields on the visible side of the Sun.
Figure~\ref{F-field_lines} shows a side view of the complex of
regions 501, 503, and their surroundings. This complex was located
at the solar disk center on 18 November.
Figure~\ref{F-field_lines}a shows the location of the null point
(slanted cross) inside this complex, the pre-eruptive U-shaped
filament F1 (red semicircular arrow), and region of bifurcation
Rb, which had well-pronounced counterparts on the solar disk
visible in the H$\alpha$ line and in soft X-rays (see Paper~I).
Figure~\ref{F-field_lines}b shows the magnetic field lines, which
pass close to the null point. According to the classification of
\inlinecite{Parnell1996}, here we are dealing with a negative
improper three-dimensional null point with the fan field lines
rooted in the N polarities, and the field lines around the spine
rooted in the S polarities. The fan is perpendicular to the spine.

The region of bifurcation Rb is rather close (but not exactly
co-spatial) to the site, where the spine field line, which leaves
the null point, enters the photosphere. The lack of coincidence in
our extrapolation, besides their possible actual difference, can
be due to: a)~the usage of the potential approximation, and b)~the
insufficient spatial resolution because of the limited number of
the spherical harmonics. There are indirect indications of a
connection between Rb and the null point:

1)~Rb firstly appeared as an isolated bright point in the
H$\alpha$ and SXI images one minute later than the U-shaped
filament started to move at 07:41~UT. The Rb region and the
filament were not connected by field lines. Their connection is
possible \textit{via} reconnection at the magnetic null point
located between them. Invoking magnetic reconnection with a
possible transformation of the null point into a current sheet is
not the only possibility here. The MHD disturbances generated by
the initial displacement of the eruptive filament can cause local
plasma heating in the vicinity of the null point by the
accumulation and dissipation of the energy of fast-mode MHD waves
or Alfv{\'e}n waves along the spine (see, \textit{e.g.}, the
review by \opencite{McLaughlin2011};
\opencite{AfanasyevUralov2012}). Such heating augments the flux of
heat and, possibly, that of non-thermal particles responsible for
the increase of the emission intensity from the chromospheric and
coronal plasmas above the photospheric base of the spine field
line.

2)~The Rb region has appeared as a small ring with a brightening
running clockwise (Paper~I) at the same time as the eruptive
filament F1 would pass the magnetic null point (this phenomenon
started at 08:08~UT). A similar situation with a considerably
larger ring was observed in an eruptive event of 1 June 2002
\cite{Meshalkina2009}. In that event, the magnetic configuration
was different, it had a funnel-like shape, and the ring was
situated above the photospheric footprint of a hemispheric
separatrix surface with a magnetic null at its top. The `magnetic
funnel' confined all the ejections, as suggested by the movies of
the event. Similar to our case, a brightening running along the
ring was observed in the 1 June 2002 event, when the eruptive
filament, being transformed into a rotating ejection, passed
through the magnetic null. The motion of this ejection through the
null must be accompanied by reconnection of the magnetic field
lines associated to the ejection with those on the separatrix
surface or nearby. A response to this process was a concurrent EUV
disturbance, which propagated along the ring-like footprint of the
separatrix surface.

\begin{figure} 
  \centerline{\includegraphics[width=0.6\textwidth]
   {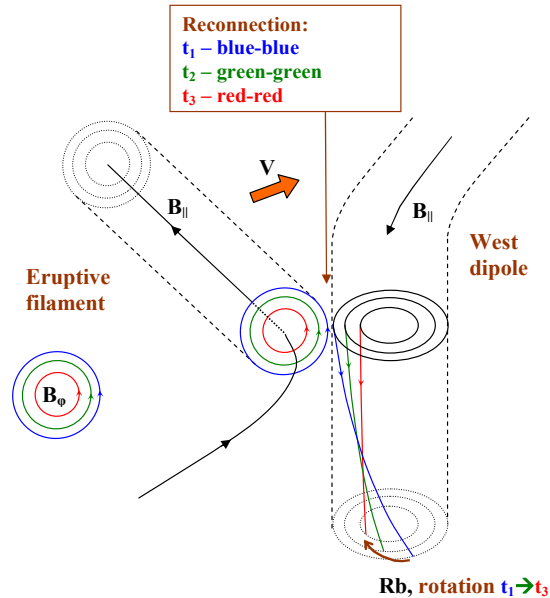}
  }
  \caption{Interaction of the eruptive flux rope (left)
and a non-uniformly twisted bundle (right) of magnetic field lines
belonging to the West dipole and rooted in Rb. Their normal
cross-sections are shown to be orthogonal in the place of
interaction. The orange thick arrow indicates the direction of
motion of the flux rope. The field lines sequentially involved in
magnetic reconnection at times $t_1$, $t_2$, and $t_3$ are denoted
with the colors presented in the top rectangle. Magnetic
reconnection leads to a concurrent brightening running clockwise
in the chromosphere and low corona above the Rb region (the brown
round arrow at the bottom). The axial magnetic fields,
$B_\parallel$, of the eruptive flux rope and the bundle correspond
to the positions of the thick blue and red loops at $t_1$ in
Figure~\ref{F-scenario}a, respectively.}
  \label{F-rotation}
  \end{figure}

In the 18 November 2003 event, the ring structure of region Rb
could be a manifestation of the photospheric base of a bundle of
magnetic field lines twisting around, or close to the spine field
line. The twist indicates the presence of the longitudinal
electric current along the spine and the inadequacy of the
potential field approximation, which we use. The twist is
necessary for the brightening running along the ring that
accompanies the concurrent magnetic reconnection of this bundle
with an arbitrary magnetic rope or eruptive filament, for example,
with a cylindrical shape. A scheme in Figure~\ref{F-rotation}
illustrates the interaction. As this Figure shows, the bundle
should be similar to a non-uniformly twisted cylindrical flux tube
to produce a running brightening. The Gold--Hoyle uniformly
twisted tube with a non-constant $\alpha$ does not satisfy this
condition, while the Lundquist magnetic cylinder with a constant
$\alpha$ satisfies it. Since Rb is located within the zone of the
negative magnetic polarity, the clockwise-running brightening
corresponds to $\alpha < 0$. In turn, this indicates that the
self-helicity of the magnetic flux around the spine is most likely
negative.

3)~In the 1 June 2002 event \cite{Meshalkina2009}, the interaction
of the ejection with the magnetic `funnel' and its passage through
the null appeared to have produced a wave disturbance in the
corona observed as an `EIT wave' rapidly expanding above the limb.
In the 18 November 2003 event, the bifurcation of the eruptive
filament was also accompanied by the appearance of a coronal wave,
whose kinematical center was located high in the corona, above Rb
(Paper~II). This was the second wave, not expected in the event.
The kinematical center of the first wave was within AR~501 and
corresponded to the impulsive acceleration stage that is typical
of impulsive eruptions \cite{Grechnev2011_I}.

\subsection{Positive Mutual Helicity of the Filament and West Dipole}
 \label{S-mutual_helicity}

Until now, the contribution to the MC magnetic helicity from the
relative position of the pre-eruptive filament and the structures
outside of AR 501 have not been taken into account. These
structures got in contact at the magnetic null point, toward which
the top of the expanding filament F1 moved.
Figure~\ref{F-field_lines} illustrates the positions of the
structures which could be involved in the eruptive process. The
values of the fluxes belonging to the West, East, North, and South
magnetic dipoles are not known \textit{a priori}. Note that each
dipole represents only a fraction of the magnetic flux of a
corresponding magnetic domain within the quadrupole configuration
in Figure~\ref{F-field_lines}. It is also not known if the
interaction between any of these dipoles and the eruptive filament
was significant. Let us estimate by linking with which of these
dipoles the pre-eruptive filament had the largest positive mutual
helicity.

The presence of the positive mutual helicity is suggested by
Figure~\ref{F-field_lines}b. The curved blue arrow in the figure
indicates the direction of the magnetic flux in the West dipole.
The red arrow indicates the direction of the axial magnetic field
in the pre-eruptive filament. Let us invoke the right-handed screw
rule and direct the screw axis along the blue arrow at its
descending spine portion; then the right-handed direction
corresponds to the direction of the red axial field in the top of
the filament. If we put the screw axis along the top of the
filament, then the result is the same; the descending part of the
blue arrow rotates clockwise. Therefore, one might expect that the
West dipole and the pre-eruptive filament constitute a
right-handed system.

\begin{figure} 
  \centerline{\includegraphics[width=\textwidth]
   {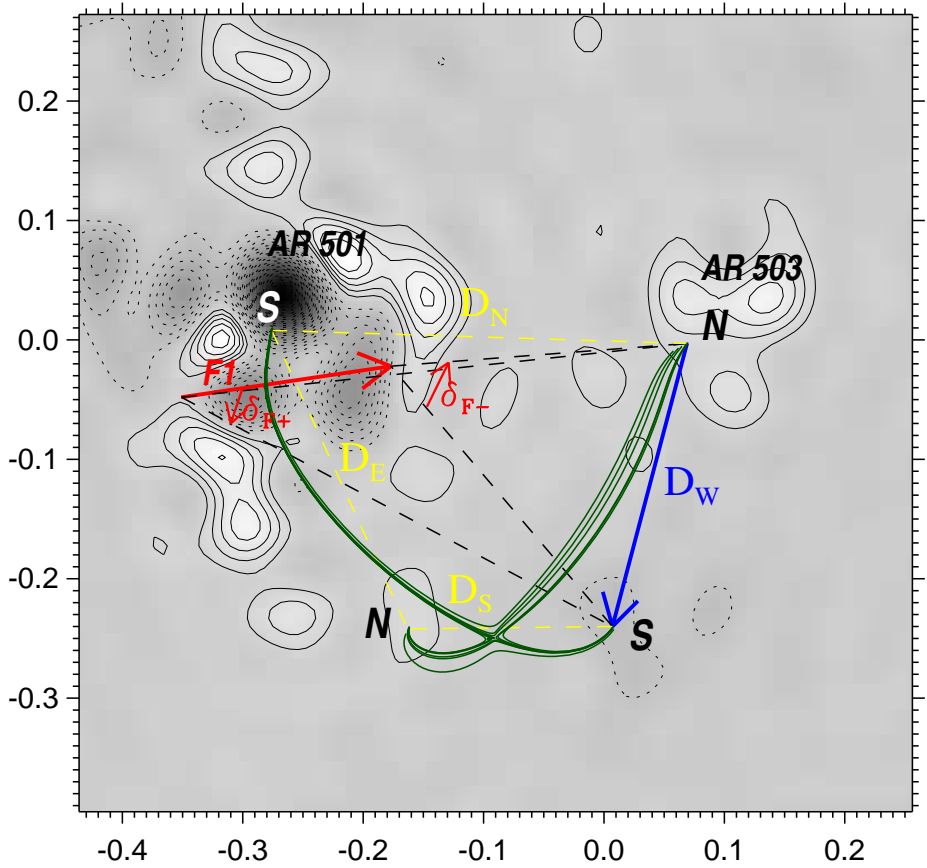}
  }
  \caption{Top view of the magnetic configuration presented in
Figure~\ref{F-field_lines}. The red vector \textbf{F1} connects
the photospheric bases of the erupted portion of the filament in
AR~501. The yellow dashed lines $D_\mathrm{N}$, $D_\mathrm{S}$,
$D_\mathrm{E}$, and the blue vector $\mathbf{D}_{W}$ connect the
photospheric bases of the North, South, East, and West magnetic
dipoles. $\delta_{\mathrm{F1}^+}$ and $\delta_{\mathrm{F1}^-}$ are
the angular spans of vector $\mathbf{D}_\mathrm{W}$ measured from
the ends of vector \textbf{F1}. The axes are in solar radii from
the solar disk center. The contour levels are symmetric relative
to zero in steps of 35~G; dotted $-(35, 70, ... 805)$~G,
continuous $+(35, 70, ... 175)$~G.}
  \label{F-top_view}
  \end{figure}

To estimate the mutual magnetic helicity,
$H_\mathrm{m}^\mathrm{mut}$, we use the technique of interior
angles \cite{Berger1998,DemoulinPariatBerger2006}. For this
purpose we need a top view of the magnetic configuration presented
in Figure~\ref{F-field_lines}b. Such a view is shown in
Figure~\ref{F-top_view}, which represents the complex of regions
501, 503, and their environment located almost at the solar disk
center when viewed from Earth. The dotted lines $D_\mathrm{N}$,
$D_\mathrm{S}$,  $D_\mathrm{E}$, and the blue vector
$\mathbf{D}_\mathrm{W}$ connect the positive and negative
footpoints of the magnetic fluxes in the North, South, East, and
West magnetic dipoles. The red vector \textbf{F1} connects the
western (negative) and eastern (positive) ends of the erupted
portion of the filament. The sign of $H_\mathrm{m}^\mathrm{mut}$
depends on the relation between the height $h_\mathrm{F1}$ of the
pre-eruptive filament and those of the magnetic loops,
$h_\mathrm{N}$, $h_\mathrm{S}$, $h_\mathrm{E}$, and
$h_\mathrm{W}$, which constitute each of the dipoles. The
estimation shown below corresponds to the real case in which the
height of the pre-eruptive filament is less than any of the other
magnetic loops. For example, the mutual magnetic helicity of
filament F1 and the West dipole $\mathbf{D}_\mathrm{W}$ is:

$H_\mathrm{m}^\mathrm{mut}(\mathrm{F1},\,\mathrm{W}) =
\Phi_{\mathrm{F1}}\Phi_\mathrm{W}(\delta_{\mathrm{F1}^+} +
\delta_{\mathrm{F1}^-})/180^{\circ} \approx
+\Phi_{\mathrm{F1}}\Phi_\mathrm{W} /10 > 0,$

\noindent where the angles $\delta$ are positive
counter-clockwise, so that $\delta_{\mathrm{F1}^+} = -33^{\circ}$
and $\delta_{\mathrm{F1}^-} = +51^{\circ}$. The flux of the axial
magnetic field in the pre-eruptive filament, $\Phi_{\mathrm{F1}}$,
is unknown. The magnetic flux of the West dipole,
$\Phi_\mathrm{W}$, will be estimated in
Section~\ref{S-double_dimming}.

The mutual helicities
$H_\mathrm{m}^\mathrm{mut}\{(\mathrm{F1},\,\mathrm{S});
(\mathrm{F1},\,\mathrm{E}); (\mathrm{F1},\,\mathrm{N})\}$ can be
estimated in a similar way. In particular,
$H_\mathrm{m}^\mathrm{mut}(\mathrm{F1},\,\mathrm{E}) < 0$. Next,
the angles between the lines $\mathrm{F1}$,
$\mathbf{D}_\mathrm{N}$, and $\mathbf{D}_\mathrm{S}$ in
Figure~\ref{F-top_view} are small; therefore,
$H_\mathrm{m}^\mathrm{mut}(\mathrm{F1},\,\mathrm{S})\gsim 0$ and
$H_\mathrm{m}^\mathrm{mut}(\mathrm{F1},\,\mathrm{N})\lsim 0$.
Small changes of their mutual orientations can change the signs of
the corresponding mutual helicities.

Thus, $H_\mathrm{m}^\mathrm{mut}(\mathrm{F1},\,\mathrm{W})$
appears to be the only solar source for a significant positive
magnetic helicity of the interplanetary magnetic cloud. If so,
then the magnetic flux of the West dipole should be involved in
the eruption of the U-shaped filament F1. This conjecture is
confirmed observationally by the development of the double
dimmings in the conjugate footpoints of the loops anchored in the
West dipole (see Section~\ref{S-double_dimming}).

\subsection{Double Dimmings in the West Dipole}
 \label{S-double_dimming}

\begin{figure} 
  \centerline{\includegraphics[width=\textwidth]
   {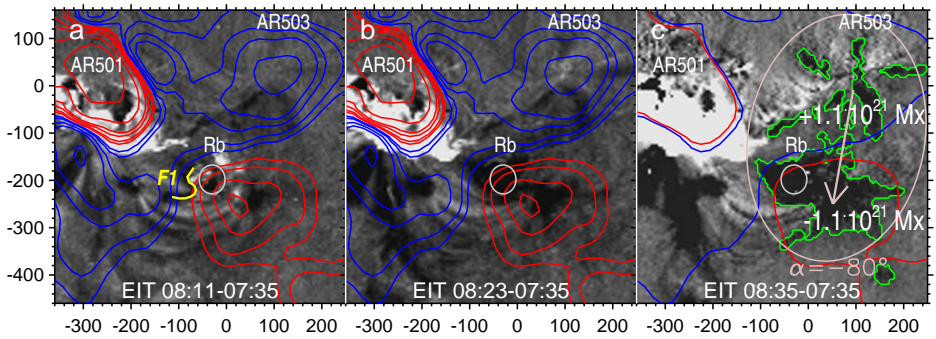}
  }
  \caption{Manifestations of a presumable eruption at the solar disk center
around 08:14~UT shown by the development of dimmings in EIT
195~\AA\ images overlaid with the contours of the coronal $B_r$
magnetogram extrapolated from an MDI magnetogram at a height of
28~Mm (blue N-polarity, red S-polarity). The contour levels are
$\pm (2.5, 5, 10, 15, 30, 60)$~G in panels (a) and (b) and $\pm
2.5$~G in panel (c). The yellow contour in panel (a) outlines the
leading edge of the eruptive filament pouncing on a coronal
obstacle above the bifurcation region Rb denoted by the small
oval. The green contours in panel (c) outline the darkest dimming
areas, which overlay a dipole outlined with a large oval. The
arrow shows the direction of the resulting magnetic field in the
dipole.}
  \label{F-dimming_mag}
  \end{figure}

Figure~\ref{F-dimming_mag} presents the development of dimming
around 08:14~UT in association with the main eruption on 18
November. The gray scale background in the panels shows three
fixed-base ratios of EIT 195~\AA\ images, $\mathrm{Ratio}_j =
\mathrm{Image}_j/\mathrm{Image}_1$. The contours outline the
positive (blue) and negative (red) modeled flux concentrations of
the coronal $B_r$ magnetogram. Here $B_r$ is the radial component
of the coronal magnetic field computed at the spherical surface
with a radius $R_\mathrm{mag} = 1.040R_{\odot}$ (\textit{i.e.}, at
a height of $h_\mathrm{mag} = 28$ Mm above the photosphere) from
the SOHO/MDI magnetogram observed at 07:59~UT.

Figure~\ref{F-dimming_mag}a presents the collision of the eruptive
filament F1 (yellow) with a high coronal structure. Their
interaction is indicated by the brightenings turning around the
bifurcation region (Rb) underneath (see Paper~I). A large dimming
South of AR~501 resulted from a preceding eruption and is beyond
our scope. Figure~\ref{F-dimming_mag}b (12 minutes later) presents
the outcome of the collision. Both the brightenings and region Rb
disappeared. Two dimming areas developed instead. A new large
dimming appeared at the previous position of the bifurcation
region, and a weaker star-like dimming developed in AR~503.
Figure~\ref{F-dimming_mag}c shows the situation observed still 12
minutes later. The green contours outline the darkest portions of
the two major West dimming regions.

The star-like dimming in AR~503 and the dimming encompassing the
former bifurcation region overlaid a dipole pointing nearly South.
The magnetic flux in this dipole probably corresponds to the
magnetic flux in a presumable eruptive structure. The value of the
magnetic flux computed within the dimming regions depends on the
height, $h_\mathrm{mag}$, to which the magnetogram is related.
This can be a photospheric magnetogram actually observed with MDI
or a coronal $B_r$ magnetogram computed from the potential field
extrapolation. To estimate the magnetic flux in the eruption, we
assume that the dipolar double dimming developed mainly due to the
stretching of a closed magnetic flux tube, connecting these areas.
One might expect that the positive and negative magnetic fluxes
within the conjugate dimming areas are equal in absolute value.
The equality should be reached at a proper height above the
photosphere. At the photospheric level, $h_\mathrm{mag} = 0$, the
computation gives a negative imbalance of the magnetic fluxes,
$\Phi^{+} = +1.8 \times 10^{21}$ Mx in the star-like dimming and
$\Phi^{-} = -3.9 \times 10^{21}$ Mx in the large South dimming.
While the height increases, the imbalance of the fluxes decreases.
It becomes zero at $h_\mathrm{mag} = 28$~Mm, where
$\Phi_\mathrm{dim} \approx \Phi^{+} \approx  |\Phi^{-}| \approx
1.1 \times 10^{21}$~Mx, and then becomes increasingly positive.
The height at which the flux is balanced is consistent with a
typical height range of the coronal emission in the 195~\AA\ line.

Considering each pole of the dipole as the centroid of the
magnetic field distribution within each dimming region, we compute
an orientation of the dipole of $170^{\circ}$ with respect to the
North or about $-80^{\circ}$ with respect to the ecliptic (in the
GSE coordinate system). The gray arrow in
Figure~\ref{F-dimming_mag}c connects the centroids. The direction
of the arrow and its ends acceptably match the blue arrow of the
West dipole in Figure~\ref{F-top_view}.

Our previous discussion provides a confirmation of the
participation of the West dipole in the formation of the
Earth-directed MC with a positive helicity. The estimated magnetic
flux, $\Phi_\mathrm{dim} \approx 1.1 \times 10^{21}$~Mx, is
adequate to that in the magnetic cloud near Earth,
$\Phi_\mathrm{MC} \approx 0.55 \times 10^{21}$~Mx, estimated by
\inlinecite{Moestl2008}. The estimated magnetic flux in the West
dipole is sufficient for the MC, while the two-fold excess could
be lost by reconnection in the interplanetary space as proposed by
\inlinecite{Moestl2008} and consistent with the fact that the MC
crossed the sector boundary of the interplanetary magnetic field.
The inclination of the MC to the ecliptic plane estimated by
different authors within a range of $\theta = -(49-87)^{\circ}$
(see \inlinecite{Moestl2008}) corresponds to the orientation of
the West dipole $\alpha \approx -80^{\circ}$ in Figures
\ref{F-top_view} and \ref{F-dimming_mag}c.

\section{Formation of the Magnetic Cloud}
 \label{S-mc_formation}

\subsection{Requirements for the MC Formation}
 \label{S-mc_requirements}

According to our analysis in Papers I, II, and IV, the MC hitting
the Earth on 20 November had the following characteristics: i)~it
was formed close to the solar disk center in the bifurcation of
the eruptive filament, ii)~it was compact and disconnected from
the Sun, and iii)~it had an atypical spheromak-like configuration.
The outcome of Section~\ref{S-large_scale_config} suggests that
the MC could be formed by the interaction between, at least, two
magnetic structures, whose mutual helicity was positive before the
eruption. These structures were the West dipole and the
pre-eruptive filament. The position of the West dipole on the
solar disk meets requirement~i).

An additional requirement follows from the total magnetic helicity
conservation $H_\mathrm{m}^{\mathrm{total}} =
H_\mathrm{m}^{\mathrm{mut}}(\mathrm{F1,\,W}) +
H_\mathrm{m}^{\mathrm{self}}(\mathrm{F1}) +
H_\mathrm{m}^{\mathrm{self}}(\mathrm{W})$, where
$H_\mathrm{m}^{\mathrm{self}}(\mathrm{F1})$ and
$H_\mathrm{m}^{\mathrm{self}}(\mathrm{W})$ are the self-helicities
of the pre-eruptive filament and the West dipole, and
$H_\mathrm{m}^{\mathrm{mut}}(\mathrm{F1,\,W})$ is their mutual
helicity. The final configuration of the eruption depends on the
sign of the total helicity. However, it is not possible to
determine the sign of $H_\mathrm{m}^{\mathrm{total}}$ from the
analysis of \inlinecite{Chandra2010} and our considerations in
Section~\ref{S-helicity_absence}. The preceding estimates of the
helicity sign were related to AR~501 only and did not contain
information about the mutual helicity between the filament in the
active region and magnetic structures outside of AR~501. The
results of Sections \ref{S-mutual_helicity} and
\ref{S-double_dimming} show that if the magnetic fluxes of the
west dipole, $\Phi_\mathrm{W}$, and the filament,
$\Phi_{\mathrm{F1}}$, are related as $\Phi_\mathrm{W} <
10\Phi_{\mathrm{F1}}$, then $H_\mathrm{m}^{\mathrm{total}} < 0$
under typical assumptions.

Thus, if the total helicity of two interacting fluxes is
transformed through reconnection into the self-helicity of a
single eruptive structure, then a right-handed MC cannot be
formed. The situation is different, if the interaction and a chain
of magnetic reconnections are followed by the formation of two
eruptive structures rather than a single one. In this way,
redistribution of the magnetic helicity is possible with almost
the full transformation of the mutual helicity of two reconnected
fluxes into the self-helicity of one of the resulting structures.
The second structure can carry away practically the whole negative
helicity. The mutual helicity of the two new eruptive structures
approaches zero as they separate from each other, and therefore
can be neglected.

\subsection{Trajectory and Mass Depletion of the Eruptive Filament}
 \label{S-trajectory}

As a next step we try to understand: i)~the possible geometry of
the interaction between the eruptive filament F1 and the magnetic
loops in the vicinity of the null point (we associate its
projection on the solar disk with the bifurcation region Rb) at
their first contact and just after the passage of the null point,
and ii)~why this interaction was followed by the visible dispersal
over the solar surface of the cool plasma, which initially
belonged to the eruptive filament F1.

\begin{figure}    
   \centerline{\includegraphics[width=\textwidth]
        {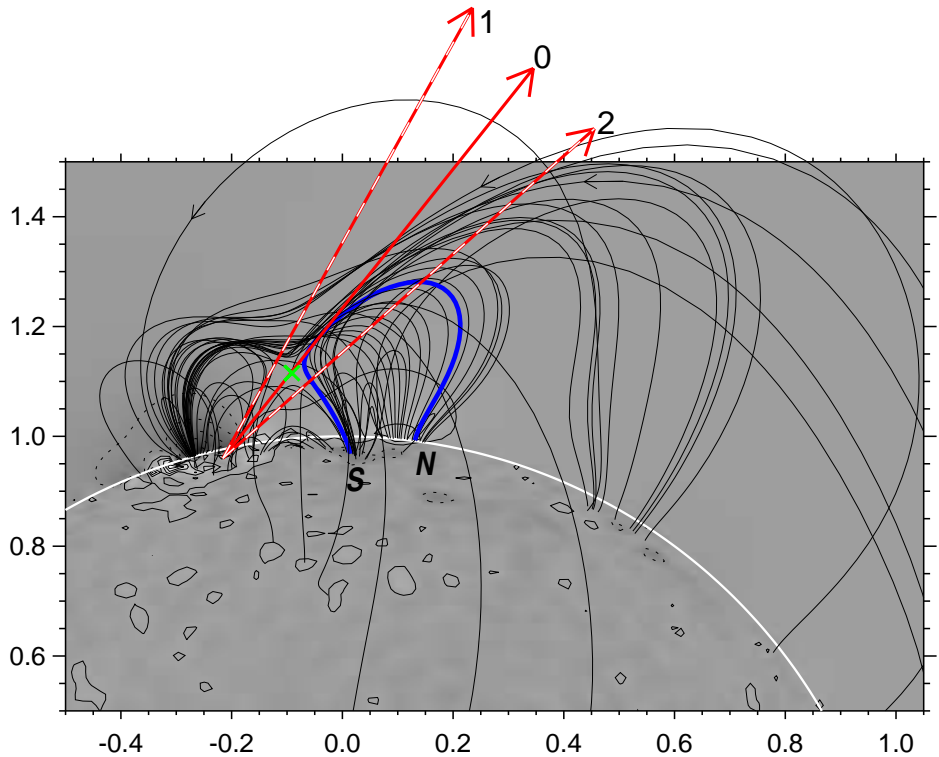}
              }
    \caption{Side view of the magnetic configuration including arrows
that indicate the trajectory of the upper part of the magnetic
flux rope replacing the erupting filament F1. The trajectory of
the rope (0) crosses a magnetic null point (green slanted cross).
The cross section of the expanding rope is limited with the red
dashed arrows 1 and 2. The thick solid curves correspond to the
magnetic field lines. The blue magnetic field line denotes the
magnetic flux from the West dipole (same as in
Figure~\ref{F-field_lines}b). The thin contours on the solar
surface correspond to the magnetic field values in steps of 34~G;
they are $-25$~G (dotted) and $+9$~G (continuous) away from the
stronger-field regions near the limb. The thick white circle
denotes the solar limb. The axes are in solar radii from the solar
disk center.
              }
 \label{F-fil_motion}
   \end{figure}

Figure~\ref{F-fil_motion} shows a side view of the magnetic
configuration, in which the main eruption occurred. For further
considerations it is convenient to replace the erupting filament
by a magnetic flux rope. The rope has a toroidal (axial) and
poloidal (azimuthal) components of the magnetic field. The
direction of motion of the middle part of the rope (the solid red
arrow marked 0) crosses the magnetic null point (the green slanted
cross). The red dashed arrows marked 1 and 2 limit the cross
section of the expanding rope. A selected magnetic field line
(thick blue) denotes a magnetic loop of the West dipole, which was
the major partner of the eruptive rope in the creation of the
spheromak.

\begin{figure} 
  \centerline{\includegraphics[width=0.8\textwidth]
   {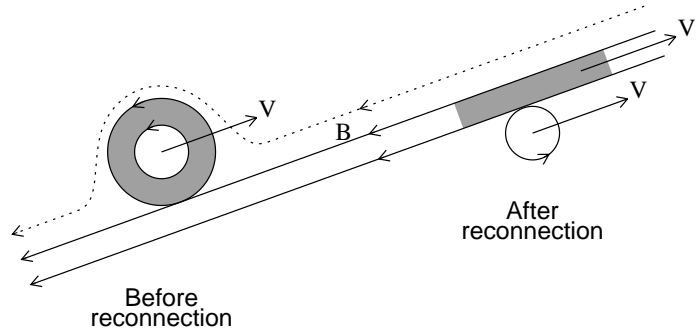}
  }
  \caption{Mass depletion in the eruptive filament moving along
the magnetic field. Left: two circular poloidal field lines
indicate the cross section of the flux rope representing the
erupting filament. Cool plasma between them (gray shading) moves
together with the rope with a velocity $V$. Right: after magnetic
reconnection. The center of the rope displaces across the outer
field $\mathbf{B}$, and cool plasma continues moving along
$\mathbf{B}$.}
  \label{F-reconnect_dispersal}
  \end{figure}

The interaction between the eruptive magnetic flux rope and
surrounding magnetic fields can result in two effects. One is the
redistribution of masses due to magnetic reconnection between the
flux rope and the outer magnetic domain. This domain contains
field lines, which reach the solar surface far from the
quadrupole. In the projection presented in
Figure~\ref{F-fil_motion}, the red solid line 0 is tangent to such
field lines. This effect is represented in
Figure~\ref{F-reconnect_dispersal} as implied in two-dimensional
reconnection models. Another effect is a kinematic linkage between
the interacting structures. This essentially three-dimensional
effect is used in a scheme presented in Figure~\ref{F-scenario}.

Figure~\ref{F-reconnect_dispersal} shows a cross section of the
top of the magnetic flux rope. This section moves along the
external magnetic field $B$ with a velocity $V$. In
Figure~\ref{F-fil_motion} this situation corresponds to the
position of the flux rope's center just after passing the magnetic
null. The two circles to the left of the figure are poloidal field
lines of the flux rope. Cool plasma enclosed between them is
denoted by the gray shading. The situation after magnetic
reconnection is shown to the right of the figure. The center of
the flux rope has shifted across $B$, while plasma (gray) has
departed from the flux rope and moves along the outer field $B$.
The scheme in Figure~\ref{F-reconnect_dispersal} is basically
similar to Figure~6 in \inlinecite{Grechnev2013_anomal}. The
apparent difference between the schemes is due to different
directions of motions of the eruptive filaments through the
magnetic null point in the quadrupolar magnetic configuration.

The 2D scheme in Figure~\ref{F-reconnect_dispersal} does not
change if we introduce an additional homogeneous magnetic field,
$B_\perp$, perpendicular to the plane of the figure without fixed
ends and consider a non-compressive plasma. During magnetic
reconnection, the frozen-in $B_\perp$ components of the total
magnetic field are mixed like pens in a box without any effect on
the process presented. However, this is not the case. The magnetic
field perpendicular to the plane of the figure is strongly
inhomogeneous being concentrated in the curved toroidal flux rope,
which continues its motion governed almost entirely by the
toroidal propelling force. Reconnection creates new
three-dimensional field lines between the footpoints of the
eruptive flux rope and those of the outer field lines involved in
the reconnection process. The outcome is as follows: i)~the
eruptive filament loses mass, ii)~the propelling toroidal force
decreases faster than without reconnection, iii)~the
disintegrating eruptive flux rope separates from cool plasma
spreading out behind it.

\subsection{Formation of the Right-Handed Spheromak}
 \label{S-scenario}

We have found that the compact magnetic cloud in the 18--20
November 2003 event could be the result of the redistribution of
the magnetic helicity after the interaction between the
left-handed eruptive filament and the West dipole $D_\mathrm{W}$.
The conclusion about the positive mutual helicity between the
pre-eruptive filament F1 and the West dipole drawn in
Section~\ref{S-large_scale_config} implies the inequality
$h_\mathrm{F1} < h_\mathrm{W}$. This means that the height above
the photosphere, $h_\mathrm{F1}$, of the red loop, which
represents filament F1 in Figure~\ref{F-field_lines}, was less
than that of the blue loop, which represents the loops belonging
to dipole $D_\mathrm{W}$. This was really the case before the
eruption. If the inequality would have been the reverse; then, the
sign of the mutual helicity would have changed. In an intermediate
case, $h_\mathrm{F1} = h_\mathrm{W}$, the mutual helicity is zero.
The conservation of the total magnetic helicity restricts the
choice of possible options for the subsequent dynamic
reconfiguration of eruptive magnetic structures. In the absence of
a clear idea about the formation mechanism of a compact
right-handed spheromak-like configuration (called henceforth the
spheromak for brevity), we use the following heuristic
consideration.

Let us consider a straight trajectory along which the centroid of
the eruptive filament moves. Arrow~0 in Figure~\ref{F-fil_motion}
corresponds to this trajectory across the magnetic null point. The
passage of the centroid through the null point (whose position is
assumed to be fixed) corresponds to the equality $h_\mathrm{F} =
h_\mathrm{W}$, where $h_\mathrm{F}$ is the height of the centroid.
This equality means that the mutual helicity between the moving
filament and the loops anchored to the West dipole,
$H_\mathrm{m}^{\mathrm{mut}}(\mathrm{F,\,W})$, becomes zero. To
keep $H_\mathrm{m}^{\mathrm{mut}}(\mathrm{F,\,W})
> 0$ until the onset of the interaction between these structures,
the trajectory of the eruptive filament should be below the null
point. This could correspond to, \textit{e.g.}, arrow~2 in
Figure~\ref{F-fil_motion}, if the condition $h_\mathrm{F1} <
h_\mathrm{W}$, which was valid before the eruption, is also valid
in the dynamic regime. In this case, the clockwise angle between
arrows 0 and 2 corresponds to a positive mutual helicity.
Similarly, arrow~1 in Figure~\ref{F-fil_motion} corresponds to a
trajectory above the null point. Some consequences of this
geometry including magnetic reconnection and mass depletion are
shown in the two-dimensional Figure~\ref{F-reconnect_dispersal}.
The counter-clockwise angle between arrows 0 and 1 corresponds to
a negative mutual helicity between the eruptive filament and the
loops of the West dipole.

\begin{figure}    
   \centerline{\includegraphics[width=\textwidth]
        {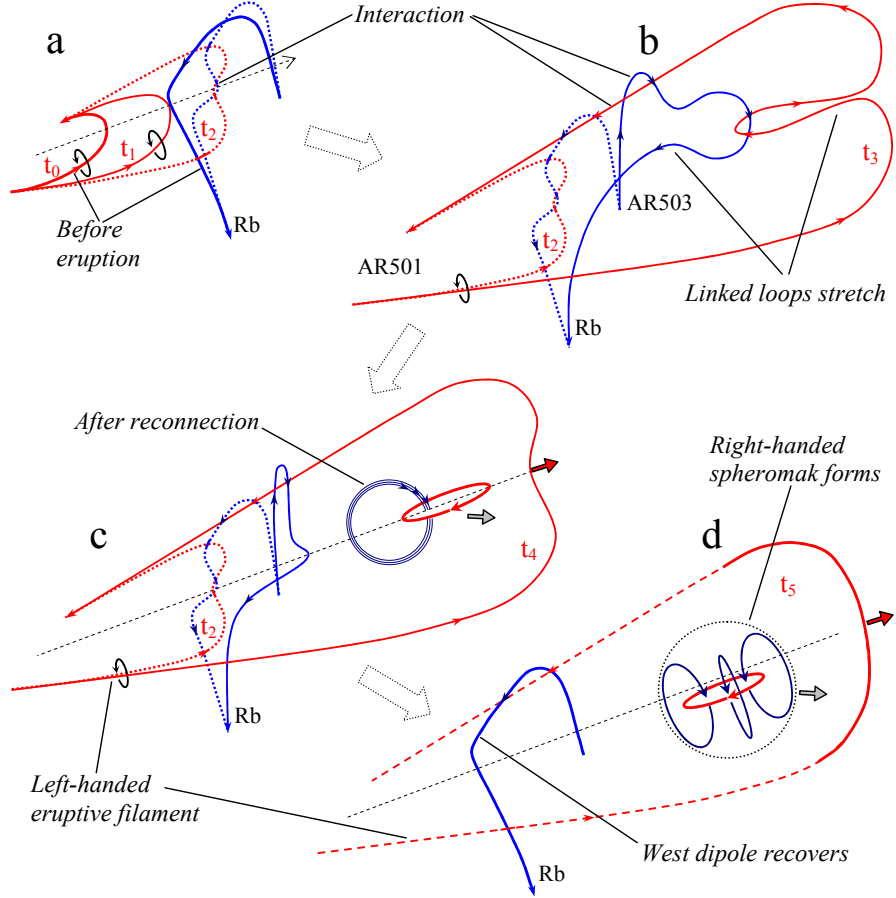}
              }
    \caption{A hypothetical development of the situation presented
in Figures \ref{F-field_lines}b and \ref{F-top_view} for the case
when the trajectory (black dashed line) of the eruptive filament
(active red loop) passes below the magnetic null point (arrow 2 in
Figure~\ref{F-fil_motion}). The blue loop is anchored to the West
dipole involved in the eruption. Rb is the bifurcation region
shown in Figures \ref{F-shape_transform}, \ref{F-field_lines}, and
\ref{F-dimming_mag}. (a)~$t_0$: denotes the time before the
eruption; the interaction of the loops starts at $t_1$. The dotted
lines at $t_2$ show the loops just after the onset of their
interaction. The expanding red loop embraces the leg of the blue
loop rooted in Rb and takes the Y-like shape (\textit{cf.}
Figure~\ref{F-shape_transform}). (b)~The stretching of the red and
blue magnetic structures results in the formation of secondary
loops which are linked. (c)~Magnetic reconnection detaches the
secondary loops from the parent ones. A closed system of red and
blue magnetic rings is formed. (d)~The blue rings separate from
each other and distribute along the red ring thus destroying
remnants of its left helicity. A right-handed spheromak develops.
The apparent intersections of the blue and red loops in panels
(a)--(d) are due to the projection effect and do not imply the
possibility of reconnection. Figure~\ref{F-suggestion} roughly
outlines the dynamic reconfiguration presented in panels (a) and
(b) of this figure, but viewed from a different angle.
              }
 \label{F-scenario}
   \end{figure}

Figure~\ref{F-scenario} presents a hypothetical 3D scheme for the
formation of a right-handed spheromak, if the trajectory is below
the null point. The relative position of the red and blue thick
solid loops in Figure~\ref{F-scenario}a is the same as in
Figure~\ref{F-field_lines}b, which shows the situation before the
eruption ($t_0$). The active red loop represents the left-handed
eruptive filament during subsequent times $t_1,...,t_5$. The
passive blue loop is anchored to the West dipole involved in the
eruption. The red and blue dotted lines represent the modified
shapes of the red and blue loops just after the onset of the
interaction. The expanding red loop embraces the leg of the blue
loop rooted in Rb and takes a Y-like shape (\textit{cf.}
Figure~\ref{F-shape_transform}). The stretching of the red and
blue magnetic structures in Figure~\ref{F-scenario}b leads to the
formation of secondary loops which are linked. Magnetic
reconnection results in the detachment in Figure~\ref{F-scenario}c
of the secondary magnetic loops from their parent loops. A closed
system of the orthogonal red and blue magnetic rings is formed.
The blue ring developed from the magnetic flux of the West dipole.
We neglect its self-helicity and represent it as a system of
non-twisted thin blue rings instead of a single thick blue ring.
The magnetic pressure separates these rings (not interconnected)
from each other, and they get distributed along the red ring.
Remnants of its negative helicity annihilate in this way
(Figure~\ref{F-scenario}d). A right-handed spheromak is formed.
The direction of its motion does not necessarily coincide with the
direction, in which the eruptive filament, which takes back its
shape, moves (the thick red arrow).

\subsection{Observational Consequences of the Proposed Schemes}
 \label{S-consequences}

The outcome of the process presented in Figure~\ref{F-scenario} is
the formation of two different eruptive structures sharing their
positive and negative magnetic helicity. The first structure is a
modification of the parent left-handed eruptive filament F1, which
recovers after the bifurcation and moves in the southwestern
sector of the solar disk (Figure~\ref{F-shape_transform}). The
second structure is the right-handed spheromak, which was created
due to the catastrophe of the filament observed as its
bifurcation.

Formally assuming that the effects presented in Figures
\ref{F-reconnect_dispersal} and \ref{F-scenario} coexisted in the
real eruptive process, we come to the following conclusion. The
bifurcation of the eruptive filament F1 was accompanied by the
decrease of its poloidal flux and mass depletion. The loss of mass
accounts for the absence of a conspicuous core in CME2, which
appeared in LASCO images in the same southwestern sector, where
the eruptive filament F1 moved. The Y-like trace of the mass,
which has not left the Sun, carries information about the
deformation of the magnetic flux rope in the vicinity of the
magnetic null point. If the magnetic cloud, which reached Earth,
would have been associated with CME2, then its magnetic helicity
should be negative rather than positive. This association seemed
to be plausible in the first studies of the 18--20 November 2003
event, but was ruled out in our Paper~II.

In our considerations, we called the `spheromak' a spheromak-like
force-free magnetic configuration. The development of a such a
configuration should be accompanied by a transformation of the
magnetic energy excess of the interacting pre-spheromak structures
into the kinetic energy of chaotic or directed plasma motions. It
is possible that some part of this kinetic energy had gone into
the formation of a wave, possibly a shock, propagating away from
the formation site of the spheromak. An indication of such a wave
is presented in Paper~II (see also
Section~\ref{S-null_and_bifurcation_region}, last paragraph).

While constructing the scheme in Figure~\ref{F-scenario}, we
pursued to have as a result the compact size and positive helicity
of the developing magnetic structure to match the MC, which
actually hit Earth. The spheromak meets these requirements.
However, our hypothetical scheme does not guarantee an earthward
direction of the spheromak. There are additional indications
supporting that the spheromak has actually reached Earth.

The formation of the spheromak in Figure~\ref{F-scenario} is due
to the interaction of different-temperature plasma structures,
\textit{i.e.}, the cool eruptive filament and the magnetic flux of
the West dipole frozen into the coronal plasma. Probably, this
circumstance determined the atypically high inhomogeneity of the
temperature distribution in the 20 November MC mentioned in
Section~\ref{S-introduction}.

The axial magnetic field of the spheromak in
Figure~\ref{F-scenario} is formed from the magnetic flux of the
West dipole. Therefore, the orientation of the axial field of the
spheromak should be close to the direction of the vector
$\mathbf{D}_\mathrm{W}$ in Figure~\ref{F-top_view}. In turn, the
inclination of $\mathbf{D}_\mathrm{W}$ to the ecliptic plane is
reasonably close to the inclination of the MC
(Section~\ref{S-double_dimming}, last paragraph).

To conclude this section, we note that the scheme of the
interaction between the red and blue loops in the case of an upper
trajectory 1 in Figure~\ref{F-fil_motion} also provides the
possibility to form a spheromak, but with a negative helicity.
Most likely, this option did not occur in this event, as supported
by the downward displacement of the central part of the eruptive
flux rope in Figure~\ref{F-reconnect_dispersal}. In the magnetic
configuration presented in Figure~\ref{F-fil_motion} this effect
works in the same direction both before the passage of the null
point by arrow 0 and after it. Thus, the lower trajectory appears
to be preferential and the only possibly one in this particular
event.

\section{Summary}
 \label{S-summary}

The scenario of the 18 November 2003 event does not correspond to
the concept of a simple eruption directly from AR~501, in which
the twist helicity of an eruptive structure or active region
determines the handedness of the interplanetary magnetic cloud.
The NLFF extrapolation of AR~501 shows an excess of negative
twist, which is opposite to the positive sign of twist in the MC.
To solve this contradiction, we have used the positive mutual
helicity between the pre-eruptive filament and the flux tubes of a
magnetic domain of the large-scale quadrupole configuration. The
interaction of these magnetic fluxes presumably occurred as the
eruptive filament passed in the neighborhood of the coronal
magnetic null point. The positive mutual helicity of these two
fluxes changed through magnetic reconnections into the positive
self-helicity of a spheromak-like structure, whose geometry and
parameters correspond to the magnetic cloud, which reached Earth.
In Paper~IV, we analyze the interplanetary disturbance responsible
for the 20 November superstorm and outline the overall scenario of
the whole event.

\begin{acks}

We thank V.~Yurchyshyn, who kindly supplied the vector magnetogram
of AR~501 observed at BBSO, and M.~Temmer for the H$\alpha$ data.
We thank the co-authors of our Papers I, II, and IV, who are not
involved in this study. We appreciate an anonymous reviewer for
valuable remarks and comments. We are grateful to the instrumental
teams of the Kanzelh{\"o}he Solar Observatory; TRACE and CORONAS-F
missions; MDI and EIT on SOHO (ESA and NASA) for the data used
here. This study was supported by the Russian Foundation of Basic
Research under grants 11-02-00757, 11-02-01079, 12-02-00008,
12-02-92692, and 12-02-00037, and the Ministry of Education and
Science of Russian Federation, projects 8407 and 14.518.11.7047.
\end{acks}

\end{article}

\end{document}